\documentclass[conference]{IEEEtran}
\IEEEoverridecommandlockouts
\usepackage{cite}
\usepackage{amsmath,amssymb,amsfonts}
\usepackage{algorithmic}
\usepackage{graphicx}
\usepackage{textcomp}
\usepackage{array}
\usepackage{hyperref}
\usepackage{subfig}
\usepackage{xcolor}
\hypersetup{
  colorlinks, linkcolor=red
}
\def\BibTeX{{\rm B\kern-.05em{\sc i\kern-.025em b}\kern-.08em
    T\kern-.1667em\lower.7ex\hbox{E}\kern-.125emX}}
\begin{document}

\title{A systematic literature review on Virtual Reality and Augmented Reality in terms of privacy, authorization and data-leaks \\
}

\author{\IEEEauthorblockN{ Parth D Patel }
\IEEEauthorblockA{\textit{Department of Engineering} \\
\textit{University of Guelph}\\
Guelph, Canada \\
parthdip@uoguelph.ca}
\and
\IEEEauthorblockN{ Prem Trivedi }
\IEEEauthorblockA{\textit{Department of Engineering} \\
\textit{University of Guelph}\\
Guelph, Canada \\
prempiyu@uoguelph.ca}
}
\maketitle

\begin{abstract}
In recent years, VR and AR has exploded into a multimillionaire market. As this emerging technology has spread to a variety of businesses and is rapidly increasing among users. It is critical to address potential privacy and security concerns that these technologies might pose. In this study, we discuss the current status of privacy and security in VR and AR. We analyse possible problems and risks. Besides, we will look in detail at a few of the major concerns issues and related security solutions for AR and VR. Additionally, as VR and AR authentication is the most thoroughly studied aspect of the problem, we concentrate on the research that has already been done in this area.
\end{abstract}


\section{Introduction}
In today's time where it's so easy to get lost in the world of the internet  \cite{c1}. To read about something online that is not even accessible to the person is what makes the internet so wonderful. The better thing is being able to experience those things without even leaving your house. This is where virtual reality comes (VR) comes in, reading about hot and humid rainforests in Africa while sitting in Canada is something that was not possible until recently. VR helps a person to experience what he read in real-time through VR devices such as VR headsets \cite{b31}. VR provides a simulated experience by tracking the person's movements and displaying a set of 3D videos to present it as a virtual world. Anything could be shown in this virtual world, ranging from true scenarios like rainforests or mountains to something that was created artificially like games. Another thing we will be discussing is Augmented reality(AR). AR is something that combines real world with the virtual world. It augments the interactive experience and represents the real world with some other computer generated content. The best example for this would be a famous game called Pokemon GO. The game asks people to find new pokemons by traveling in the real world.  \\
Although they both have their benefits, there also come some security threats that cannot be ignored. Both cases' biggest and most common threat would be a user's privacy. In case of VR a person's movement can be tracked whereas in AR a person's location could be tracked as well as permission for access to their could be compromised. It possesses a greater threat for AR as it is easily accessible in different devices such as mobiles and tablets. On the other hand, to access VR one would need a VR headset. Another threat that would be common for both is content displayed. As in both cases content is provided through a third party it could be hacked and changed. Thus providing different content than originally intended. Something that could be improved in both cases would be the authentication process. In both cases, a password only known to the user is used as authentication but it could be improved. VR could use the behavioral pattern such as walking and hand movement to authenticate the user as it can not be copied. AR authentication could be improved based on the devices that are being used such as mobile or tablet could use biometric authentication whereas a PC could use face recognition.\\
As per bailenson \cite{b19} VR can track movements of the user. This nonverbal action can be used for different means, such as identifying what kinds of advertisements are relevant to the user. This could be a huge risk as this could be also be used to impersonate someone. In the case of AR the input and output both could be compromised. As stated by Roesner ankohno \cite{b16}, raw videos are provided to third party which in turn creates an augmented output. The input needs to be verified as it could compromise others privacy also the user could show something that was not intended upon, resulting in privacy leaks. Similarly, the output provided by the third party should also be trusted and verified. A very good example on how AR and VR is potentially used to manipulate people was published in \cite{b11} explaining how a popular game like Pokemon GO which uses AR uses user data to manipulate the foot traffic and also provides the same data to VR industry in order for them to improve their Virtual World.
\\
This paper focuses on analyzing more of such threats and attacks that affect the security of AR and VR and how the data leaks could be used to manipulate a person's action. We will also discuss potential solutions that could be applied or whether no solution could be found. 

\subsection{Prior Research}
To the best of our knowledge, there appear to be relatively few Systematic Literature Reviews on the application of AI/VI to the problem of cyber security (SLRs). Viswanathan Karthik \cite{b2} recently published a comprehensive study on security considerations in VI, and Allen, Paul G. Allen recently published a survey paper on security and privacy in AI\cite{b9}. The authors of both studies highlight the challenges and problems associated with the use of security services in centralised architecture in various application domains and provide a comprehensive review of current security methods in AI/VI for such security service applications in areas of authentication, confidentiality, privacy, access control, and data provenance. This study, in our opinion, provides a fundamental understanding of the cutting-edge technologies AR and VR. Furthermore, after learning about AR/VR, the research will concentrate on the security considerations in AR/VR. Furthermore, there are a few additional studies in which specific technologies, such as Google Glass in VR and how to make Glass more safe in terms of data and privacy, are examined.\\

Most of the studies are from 2022 and the oldest one referred here is from 2015, thus all the studies discussed are recent \cite{c2}. This also specifies that most of the studies regarding VR and AR were conducted in recent years. Although the attacks and threats identified are pretty generic showing that VR and AR are still in their early phases and very vulnerable. Some attacks are repeated in studies and more common, such keywords are identified after analyzing multiple papers. This keywords are “Unauthorized access”, “Human Joystick”, “Brute force”\cite{b5}. As this attacks are more common and can not be tracked easily, they could happen often until some steps are taken to prevent this. Another vulnerability are the devices that are used to access VR and AR. According to case study \cite{b4} the attackers could access the videos streamed in VR or the input given to AR. Thus a different output could be provided. In the case of AR the augmentation performed in live feed could give out different results than intended. Similarly, in the case of VR if the image fed in the console could be hacked and a different output could be shown. All such attacks will also be studied in this review. 
\subsection{Research Goals}
The goal of this study is to examine existing research on the burgeoning topics of AR and VR. This document was written and summarized with security issues of AR and VR in mind. To concentrate the effort, we established three research questions, as indicated in the \hyperref[Table1]{table I}.

\begin{table}[htbp]
\caption{Research Questions}
\label{Table1}
\setlength{\tabcolsep}{10pt} 
\renewcommand{\arraystretch}{2} 
\begin{tabular}{| p{14em} | p{3.3cm} |}
\hline
\textbf{Research Question (RQ)} & \textbf{Discussion}\\ 
\hline
RQ1: What are the most recent AR and VR security applications? & A survey of the most recent practical applications will aid in comprehending the entire scope of AR and VR technology's influence on cyber security. \\
\hline
RQ2: How are augmented reality and virtual reality being utilized to enhance cyber security ? & Every firm needs a safe cyber security infrastructure today. Through visualization, AR and VR technologies may considerably assist IT and security personnel in developing a safe and resilient security architecture. Aside from that, AR/VR technology may assist companies in developing a proactive incident response strategy.\\
\hline
RQ3: What options/methodology are there for managing security solutions in augmented reality and virtual reality ? & The most important data given by AI/VI is the concealment of personal and secret data. Encrypting the data before releasing it can provide security to this operation. In this part, similarly we will look at comparable methodologies.\\
\hline
\end{tabular}
\end{table}

\subsection{Contribution and layout}
This SLR supplements previous research by providing the following contributions for people interested in AI/VI and cyber security to advance their work:\\
- To early 2018, we identified 80 primary studies relevant to AI and VR in cyber security. This list of studies can be used by other scholars to advance their work in this topic.\\
- We then choose 16 primary studies that match our quality evaluation criteria. These studies can serve as useful standards for comparison with related research. We do a thorough evaluation of the data included in the subset of 16 studies and provide the findings in order to represent the research, thoughts, and considerations in the disciplines of AI, VR in terms of cyber security.\\
- We make statements and provide recommendations to help with future work in this area.

The following is how this paper is organised: Section 2 discusses the strategies used to select primary studies for analysis in a systematic manner. Section 3 summarises the findings of all of the primary studies that were chosen. Section 4 addresses the findings in relation to the earlier mentioned study topics. Section 5 finishes the study and makes some recommendations for further research.

\section{Research methodology}

We used the SLR to answer study questions, following the guidelines presented by Paul J. Taylor and tooska dargahi \cite{b20}. We attempted to progress through the review of papers that had been shorted out. Moreover, we will conduct and report steps in iterations to allow for a full examination of the SLR.
\subsection{Selection of primary studies}
The initial research criteria was to find papers by using the basic keywords like “Virtual reality”, “Augmented reality”, “security”,”authentication” and “privacy”. To extract the necessary articles for our research, we attempted to search for papers in various search strings using boolean operators. The search terms were:\\

(“Virtual Reality” OR “Augmented reality” ) AND (“security” OR “authentication” OR “privacy”)
\\

This enabled us to present a variety of papers, some of which were relevant and others of which were not; the same keywords were utilized on distinct sources:\\

- IEEE Xplore\\
- Google scholar\\
- UOG library\\

	Depending on the search platform, the searches were conducted against the title, keywords, or abstract. The searches were carried out on October 5, 2022, and we processed all research published up to that point. The results of these searches were filtered using the inclusion/exclusion criteria described in Section 2.2 and will consider only those papers whose inclusion matches with the searched out paper.

\subsection{Inclusion and exclusion criteria}
Studies included in this SLR must provide empirical findings and may include case studies, AI and VI security problems, or reflections on existing security procedures using AI and VI. They must undergo peer review and be written in English. Any results from Google Scholar, IEEE Xplore, or ScienceDirect will be reviewed for conformity with these criteria, as the above mentioned platform has the chances to provide lower-quality publications or studies that are too old to be included in the review. This SLR will only feature the most recent version of a study. \hyperref[Table2]{Table II} shows the important inclusion and exclusion criteria.

\begin{table}[htbp]
\caption{Research Questions}
\label{Table2}
\setlength{\tabcolsep}{10pt} 
\renewcommand{\arraystretch}{2.5} 
\begin{tabular}{| p{14em} | p{3.3cm} |}
\hline
\textbf{Criteria for inclusion} & \textbf{Criteria for exclusion}\\ 
\hline
The report must provide true information concerning AI and VI, as well as the role of security. & Paper focusing on data tracking in AI and VI. Design and Strategies in AI and VI.  \\
\hline
Paper should be concerned more with terms related to confidentiality, security, authentication, etc. & Grey literature such as blogs and government documents.\\
\hline
The article must be a peer-reviewed publication from a conference proceeding or journal. & Non-English Papers.\\
\hline
\end{tabular}
\end{table}

\subsection{Selection results}
Initially, an estimated number of research publications discovered was roughly 1600 papers. Furthermore, following sophisticated screening, the number of relevant publications was decreased to 77. After a quick scan, it was discovered that around half of these articles were either vaguely linked to the issue or not at all relevant. Another significant finding was the lack of research articles focusing on both AR and VR. All of the investigations concentrated solely on one of them. After briefly skimming over the text and finding all of the significant keywords. The research where these keywords were most commonly used and briefly described were picked, and the total number of studies where the findings were relevant amounted to 16.

\subsection{Quality assessment}
The quality of primary research was evaluated using the guidelines established by Kitchenham and Charters [20]. This allowed for an evaluation of the articles' relevance to the research issues, taking into account any indicators of study bias and the quality of experimental results. The assessment procedure was based on that of Hosseini et al. [21]. To determine their efficiency, five randomly selected articles were submitted to the following quality evaluation process:

Stage 1: \textbf{Artificial Intelligence and Virtual Intelligence.} The article must be well-commented and focused on AI and VI or the security implications of AI/VI technology to a specific problem. 
\\Stage 2: \textbf{Background.} The study aims and outcomes must be contextualized sufficiently. This will allow for a proper interpretation of the findings.
\\Stage 3: \textbf{AI/VI application}The study must include enough information to provide an accurate explanation of how the AI/VI was applied to a specific situation, which will aid in addressing research questions RQ1 and RQ2.
\\Stage 4: \textbf{Security and Privacy context} In order to aid in addressing RQ3, the paper must explain the security and privacy concerns.
\\Stage 5: \textbf{Data collection.} To assess accuracy, details on how the data was collected, measured, and reported must be provided.
\\

This quality rating criteria was then applied to all additional primary studies that were found \cite{c5,c6}.

\subsection{Data extraction}

All articles that passed the quality evaluation had their data extracted to determine the completeness of the data and the accuracy of the information included within the papers. The data extraction procedure was tested on a few studies initially before being expanded to encompass the whole collection of studies that passed the quality evaluation step. The following categories were assigned to the data: \\

\textbf{Context data:} Information on the study's purpose.\\
\textbf{Qualitative data:} The writers' findings and conclusions that are offered.\\ 
\textbf{Quantitative data:} Collecting and analyzing data from the study and generalizing the results. \\

\begin{figure}[htbp]
\centering
\includegraphics[width=9cm,height=9cm]{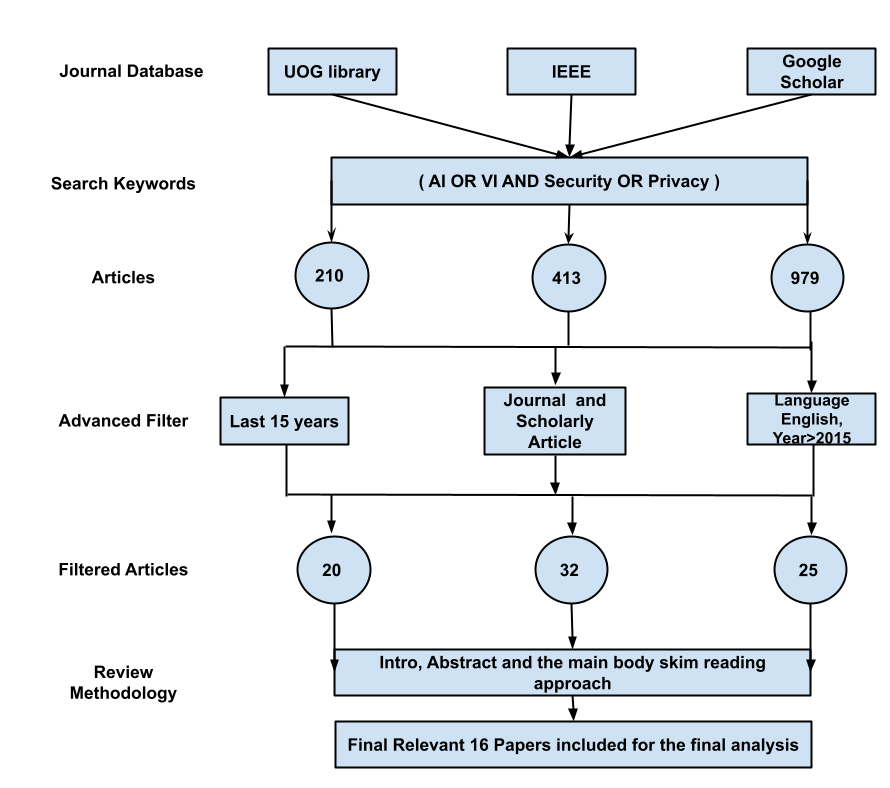}
\label{methodology}
\mbox{}
\caption{Methodology Process}
\end{figure}

\hyperref[methodology]{Figure1} depicts the process of papers selected in our research. It shows our database journal platform for finding papers using the boolean expressions to get the relevant papers. After applying advanced filter, around 80 papers are remaining that are relevant to our study. Reviewing further in details, final 16 papers were selected for this study.

\subsection{Data analysis}
We collected the data held under the qualitative and quantitative data categories to satisfy the goal of addressing the study questions. In addition, We go into details of the papers regarding authentication vulnerabilities that affect AI/VI.\\

\subsubsection{Publication over time}
The AI concept was first coined in year 1956 whereas VI was invented in 1976. Papers regarding AI/VI in terms of security, authentication or privacy were mostly found after the 20th century. This may highlight the newness of the ideas concerning cyber security applications for AI/VI .\\

\begin{figure}[htbp]
\centering
\includegraphics[width=9cm,height=7cm]{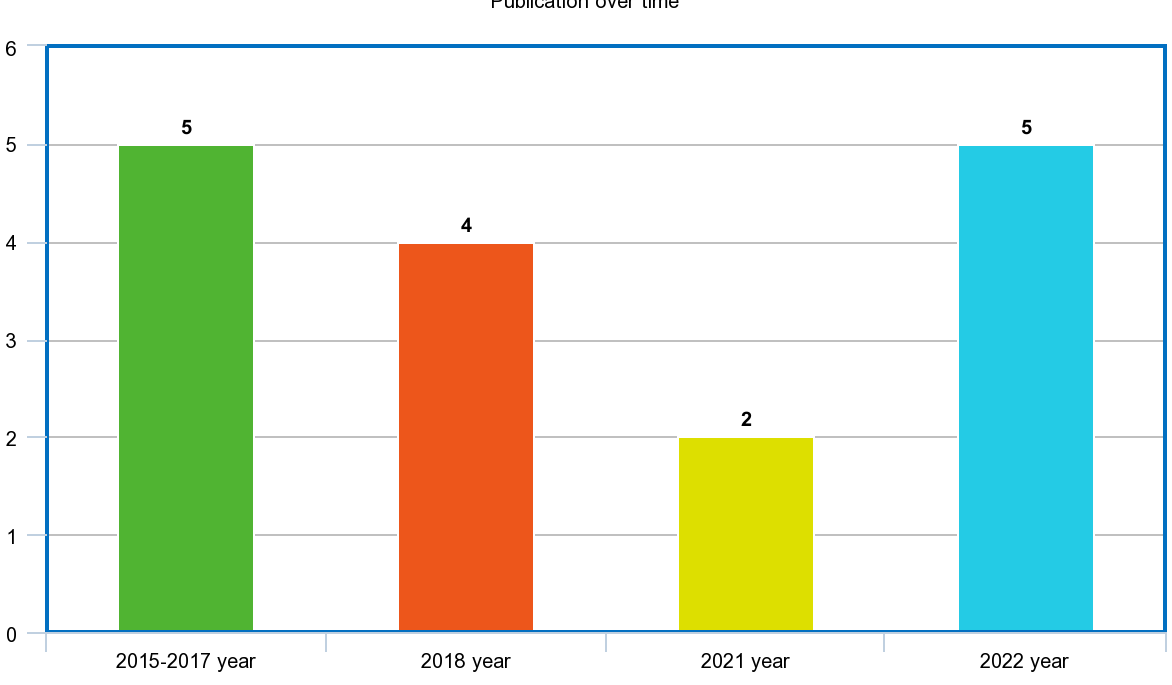}
\label{publication}
\mbox{}
\caption{The number of primary studies that have been published over time}
\end{figure}

Initially, an estimated number of research publications discovered was roughly 80 papers. After a quick scan, it was discovered that around half of these articles were either vaguely linked to the issue or not at all relevant. Another significant finding was the lack of research articles focusing on both AR and VR. We can see from the final 16 publications that there was little study done between 2015 and 2017. Furthermore, we can see that the graph of publishing has not significantly improved, indicating that researchers may be suffering, but this year has witnessed an explosion in research, with the largest number of papers published in this year, as seen in \hyperref[publication]{Figure2}.

\subsubsection{Significant keywords counts}
An analysis of keywords was done across all 16 studies in order to obtained generalize themes among the studies. As we can observe in the \hyperref[Table3]{Table III} quantitative and qualitative analysis allows us to analyse privacy problems, with authentication being the most commonly discussed topic \cite{c3,c4}.
\section{Finding}
\begin{figure}[htbp]
\centering
\includegraphics[width=9cm,height=7cm]{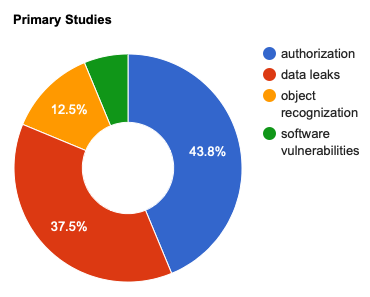}
\label{chart of theme of primary studies}
\mbox{}
\label{piechart}
\caption{The number of primary studies that have been published over time}
\end{figure}

All the primary research papers were read in full and relevant analysis is provided in \hyperref[Table3]{Table 3}. The table provides a view on what security threats are faced while using AR and VR. The major threats to these technologies were identified and then the studies were grouped in separate sections. Among these sections further subsections were classified to simplify the data. All the studies with authorization and authentication were grouped together in one section. The next section comprised of all the studies related to data storage, data leaks and data manipulation. The remaining categories are standalone and not sub-categorized which are object recognition and software vulnerabilities.
\\
It was found out that the major issue faced by this technologies is user authorization comprising of 43.8 \% of the total studies. The second group can be classified regarding data leaks and data manipulation resulting in total of 37.5\% of research papers.  The studies related to object recognition make a total of 12.5\% and the last section of software vulnerabilities makes a total of 6.3\% of the total studies. \hyperref[piechart]{Figure 3} depicts the graphical representation in form of pie-chart of above calculation.

\begin{table}[htbp]
\caption{The key research main results and topics}
\label{Table3}
\setlength{\tabcolsep}{10pt} 
\renewcommand{\arraystretch}{2.5} 
\begin{tabular}{|p{1cm}| p{14em} | p{1.5cm}|}
\hline
\textbf{Primary Study} & \textbf{Key Qualitative \& Quantitative Data Reported}& \textbf{Types of Security Applications}\\ 
\hline
\cite{p1} & Using the devices OS(most probably mobile phone) to recognize and authorize a user or an other object as desired. The device used for motion capture in the study is Microsoft Kinect and the application used for recognition of image is Privacy Googles. & Object recognition   \\
\hline
\cite{p2} & Identifying the threat in industrial AR such as unauthorized read access and building an AR edge computing architecture using edge servers to prevent such threats. & Unauthorized access\\
\hline
\cite{p3} & Object localization and object recognition in AR with the help of radio based object recognition. The device used here could be made by using localized equipment available in factories such as sensors of the mobile devices and a field technician. & Object recognition
\\
\hline
\cite{p4} & Preventing the tracking of non verbal data in VR by implementing government policies and self regulation from the companies. & Non verbal data leaks. 
\\
\hline
\cite{p5} & Understanding the privacy law and regulations in place by government  and understanding the framework and limitations with those frameworks as location leaks and information leaks through AR. & Privacy issues 
\\
\hline
\cite{p6} & Finding VR risks such as data interference and manipulation, analysing privacy policies and conducting public surveys to identify their knowledge about VR threats and privacy policies. & Privacy issues 
\\
\hline
\cite{p7} & Proposing the convergence of Social Network and VR by making digital avatars and how they could lead to informational, physical and associational privacy threats which could only be prevented by legal policies and laws. & Data leaks 
\\
\hline
\cite{p8} & Constructing an architecture for user authentication in VR such as 3D password, pattern lock and PIN. The hardware used is Leap Motion and Oculus Rift DK2 & user authentication 
\\
\hline
\cite{p9} & A study conducted between 5 people regarding their perception about VR’s data collection, its’s data collection methods and what kind of data is collected. Moreover which VR system is more trusted by the users was also discussed. & Data storage 
\\
\hline
\end{tabular}
\end{table}

\begin{table}[htbp]
\setlength{\tabcolsep}{10pt} 
\renewcommand{\arraystretch}{2.5} 
\begin{tabular}{|p{1cm}| p{14em} | p{1.5cm}|}
\hline
\cite{p10} & Checking the current user authentication protocols in AR and VR Head mounted displays, identifying their vulnerabilities and proposing a solution based on Zero - Trust algorithm(ZeTA). & User authentication 
\\
\hline
\cite{p11} & Testing popular AR and VR headmounts such as oculus Quest, Hololens, Google Glass, Valve Index and others to find out all of them have common vulnerabilities and exposure(CVEs). They also have other drawbacks such as no multi factor authentication and flawed privacy policy. & Vulnerable software 
\\
\hline
\cite{p12} & Understanding the AR output risk due to buggy and malicious third party software leading to obscuring of user’s view of the real world. Thus, proposing a solution by introducing their own AR platform Arya, which designs the output policy module.  & AR output 
\\
\hline
\cite{p13} & Constructing an access control framework called Privacy Manager. The aim is to restrict the users access to AR application in certain environments where there is a risk of private data leaks. The framework creates minimum system overhead for the mobile device while providing location awareness. & Access authorization 
\\
\hline
\cite{p14} & Analyzing different authentication methods in metaverse such as Information-Based authentication, Biometric authentication, Multi- Model authentication, Gaze- Based authentication and discussing there advantages and disadvantages. & User authentication 
\\
\hline
\cite{p15} & A study on VR authentication through Task driven biometric authorization such as hand gestures and trajectory based authentication providing results based on 135 samples. & Authentication 
\\
\hline
\cite{p16} & An experiment for eye based biometric authorization for VR application based on C++/OpenGl by implementing it on Valve’s open VR. The sample taken was in 60 Hz with non uniform time intervals. & User authorization  
\\
\hline
\end{tabular}
\end{table}

\section{Discussion}
After the initial research, some of the keywords that were found relevant are “Privacy” \cite{b5}, “Authentication” \cite{b2},  “Data leaks”. All three of these keywords are found to be most vulnerable in AR and VR security. As we continued using these keywords, we also discovered several unrelated publications that used the same nomenclature but had distinct study motivations and conclusions\cite{b17} \cite{b18}  \cite{c7} \cite{c8}.
\\

\textbf{RQ1: What are the most recent AR and VR security applications?}
\\Considering the use AR and VR in security aspects for real world, there are very limited applications that could be found. One of the application for AR that’s recent but a very prime example for real life security is using AR devices for surveillance. The problems with regular surveillance is that the image available is in 2D and needs a good amount of camera for a large floor plan. The other major drawback is the competence of the operator. As there is always a risk of error with human supervision, the AR devices which could be used for face identification or object identification as well as could also identify the environment and with the use of proper hardware could also keep view of a wide surface area in 3D.\\
One such framework is discussed in \cite{b26}, where the operator could use AR technology to provide a 3D interface with the help of Virtual and Augmented reality. This is done by using a wide array of sensors for video and audio surveillance. The model works by initially creating a sample 3D model of the environment. This model could be made complex or could be kept simple based on the operator’s requirement.\\ 
Once the 3D model of the environment is created successfully the model than uses several service discovery protocols to find all the camera in the environment create a mesh network. For each camera found in the real world the model would then represent a virtual representation of the cameras in 3D format for operator where the operator could access the whole floor plan from a single point of view. The 3D world represented to the operator is connected to the real world image thus providing an easier access for operator to keep surveillance over a wide area easily.\\
One other implementation of VR for real world physical security would firearm training simulation\cite{b27}. It is not so much of a surprise that firearm training in real world could not be done daily due to high cost of firearms. Thus the training itself is short and spaced over time. Moreover, the occurrence of surprise attacker can not be simulated in real world gun ranges thus not providing proper training. Although this could be improved with the use of VR simulation.
This approach allows more effective training as multiple aspects could be introduced in a virtual world and can be controlled like pedestrians in a street, surprise occurrence of vigilante’s other aspects, providing a proper real life simulation for the trainee. This results in professional training with a higher emphasis on daily training while keeping the cost of training to a minimum. \\

\textbf{RQ2: How are augmented reality and virtual reality being utilized to enhance cyber security ?}\\
It is clear that with current scenario of world the importance IT department in any company has significantly increased. With the introduction of IT department in almost every field the risk of it being hacked also increases creating a major problem. For employees that are not well educated with cyberattacks it becomes an alarming problem as to simple cyber attacks like phising could lead to a huge data leak. To prevent such scenarios, it is necessary for any company to have employees that are well trained regarding cyber attacks.\\
\mbox{}
Training through AR/VR is being widely used and seems to be more efficient as it gives a virtual depiction as well as provides audio to support it’s curriculum. As discussed in \cite{b28}, AR is being used in a wide array of activities like Order breakdown, Warehouse operations, Quality control and much more. Another benefit is that it could be accessed through a large number of devices such as Tablets, smartphones, data goggles providing people with visualization of decision making simulation and also providing the effects of the decisions. 
Thus, it is also possible to provide cybersecurity training through AR and VR devices. All we need is a proper curriculum. Considering multi generational businesses and a wide range of age in different businesses, the virtual depiction could help to provide a better understanding of the significance of cyber security.\\
\mbox{}
Another major application of AR is for the visualization of security data for cybernetic systems \cite{b29}. The study itself was conducted on students studying information security. The study focused on two factors which were accuracy and speed. As a result, the operator's contact with the data collected by information security systems should improve in terms of performance indicators (accuracy and speed). It was concluded that this approach should be applied for large volumes of data which have a greater significance towards accuracy and speed. \\

\textbf {RQ3: What options/methodology are there for managing security solutions in augmented reality and virtual reality?}\\
\mbox{}
The benefits of employing augmented reality tools are abundantly clear. Wearables enable workers to complete tasks more quickly, adaptably, and safely. They demonstrably make less mistakes and are happier employees. But as the name implies, smart glasses only function by gathering and analysing data. Therefore, a fundamental concern for any businesses employing AR technology is data security.\\
The best way to prevent data leaks is user authentication. Although, in the case of AR/VR the authentication could be done by hand gestures or body movements or by entering patterns and PIN which could easily be seen by a bystander or could also be recorded by someone else which could lead to someone else accessing the data. Thus, as discussed in \cite{p10}, the Zero – trust algorithm(ZeTA) could seem to prove a good fit for such cases. ZeTA is a knowledge-based authentication technique, which means that, like text passwords, the user must memorise a secret.\\
\mbox{}
The study itself used two people at a time in a single room to analyze whether one could analyze the second person could analyze the first person and vice versa.  In the experiment person 1 was authenticated three times while person 2 observed and then the positions were changed. Finally both were asked to answer questions in a survey.\\
\mbox{}
Another good methodology for authentication is using eye based biometric authorization in VR \cite{b16} as it is not possible for an outsider to track eye movements. The projet was modified in C++/OpenGl and Valve’s Open VR API was used with any VR HMD. The authentication process started with data collection then eye movement classification after which feature extraction is performed and then they try to match score calculations before taking a decision. The experiment itself was able to obtain 60 Hz signals with non- uniform time intervals but it was stated that the past experiments achieved 250 Hz signals and should be improved upon. Thus considering the current VR application, the eye based biometric authorization could be taken as a good example.

\section{Future research directions of AI/VI security}

As discussed above that AR and VR could be implemented in a wide range of fields like training to surveillance. This applications are a proof that although they are a fairly new technology but are widely applicable in the new digital world \cite{c9,c10}. One such application which takes this to another level completely changing the scenario of current world is Smart Cities.\\

A Smart \cite{b28} is a technologically advanced urban area that uses Information and communication technologies(ICT) with different kind of sensors for collecting specific type of data. This data is then used to increase the operational efficiency of the city and understand the requirements for public welfare. With the rapid advancement of technology in current era, it has become crucial to understand the dynamic infrastructure of a urban area and get certain data like air pollution, energy consumption, video surveillance and other aspects. For such analyses and data collection AR can be used to enhance the process \cite{b29}.\\

\textbf{Integration of AR and IoT for smart cities.} A distinctive immersion into Internet of Things (IoT) applications is provided by the integration of AR into smart cities. According to recent research, this can serve as a framework for an interactive demonstration of how public services like street control, video surveillance, solid waste collection, and parking management can be carried out and managed from a single platform, improving the safety, cleanliness, and livability of cities. The main purpose of a smart city is to connect everything together. This gives the users a easy and hassle free access to the all the data of city. As this provides an ease of access but also creates new risks and threats in the domain of cyber security. Users' and organisations' careless behaviour in smart cities can expose the entire city to cybercrime risk. This challenges and risk are duly noted and explained in \cite{b30}. It states that due to high reliance of smart city on ICT, several security threats like cyber attacks and leakage of private information could create a negative impact on the quality of living.\\

A review of such threats and risks was done in \cite{b27}, which states that the application of AR for smart cities can be categorized into five main classifications, including tourism, system monitoring, system management, education and instruction, and mobility. It shows the cross connection of this fields can cause major cyber attack threats. As every field and information would be interconnected then one vulnerability could lead to a major cyber attack. The study itself discusses how each area has its own pros and cons but also states that although there has been several studies regarding the attacks and proposes the solution but most of the previous work only attempts to prove concepts rather than providing with a analytical approach. Thus it could be stated that in the future we would need more research on  practical and analytical solution as the idea of smart cities keep becoming a reality.

\section{Conclusion}
This research has identified how AR and VR are widely used and are being applied to various sectors in real life. Although it seems that the technology helps in simplifying the daily needs and provides more entertainment with the help of immersive games and videos, they bring a huge drawback as the technology is still in it’s early phases. The research itself identifies some major security issues in the technology itself such as Authorization problems, data leaks, data manipulation, vulnerable software and unauthorized access. The best solution to such problems is to implement an adaptable framework that lets domain administrators and application developers implement access control policies to families of mobile applications. Implementing some policies and laws for preventing data leaks. One more solution is to create awareness among general public on how AR and VR work and what policies are implemented by the companies providing Hardware for such applications.\\
Apart from this several AR and VR security applications in real world as well as how AR and VR can be used to spread awareness regarding cyber security. The best way to apply this technologies in real world is to provide training in various sectors such as factories, firearms trainings, sales coaching and other areas. Considering the awareness regarding cyber security, a training module could be developed which provides virtual course on how the cyber attacks work and how they can be prevented. This could be most helpful in the current scenario as the society is going towards becoming a technologically advanced society.\\
Another big aspect is the application of AR in integration with IoT in smart cities. To simplify the concept this integration helps in connecting all the data collection methods and giving an easy access to all that data in a single place. Applying AR for data analysation provides a more immersive experience towards services like maintenance and installation of various infrastructures. Depending on the maintenance methods, a population of wireless nodes installed in a smart city setting can sense various factors thanks to the IoT infrastructure. Central data transport utilises any type of backbone city network. Additionally, a wireless infrastructure is provided to give city residents' and maintenance employees' smartphones network access via numerous access points (also known as WiFi APs or base stations) dispersed across the city. With the help of these components, it is possible to portray the sensors and actuators in a smart city as services that can be quickly accessible through graphical user interfaces based on actual images, without the need for a continuous connection to the backbone network.\\
Thus concluding that the AR and VR technologies have taken its roots in the industry providing a number of uses, it’s still vulnerable to many threats. To implement this technologies on a larger scale we still need to research this problems and provide a analytical solution to implement them bigger real life projects like smart cities. 

\textbf{Declarations of interest}
\\ 
\mbox{}
There is no conflict of interest.
\\

\textbf{Acknowledgement} 
\\
\mbox{}
None.

\mbox{}
\textbf{Primary Studies} 
\\

\vspace{12pt}
\color{red}

\end{document}